## Title

Multilevel latent class (MLC) modelling of healthcare provider causal effects on patient outcomes: Evaluation via simulation

## Authors

*Wendy J Harrison[1,2], Paul D Baxter[2], Mark S Gilthorpe[1,2,3]

[1]Leeds Institute for Data Analytics, University of Leeds, Leeds, LS2 9NL, UK; [2]School of Medicine, University of Leeds, Leeds, LS2 9LU, UK; [3]The Alan Turing Institute, London, UK.

*Corresponding Author: Wendy J Harrison. Leeds Institute for Data Analytics, Level 11 Worsley Building, University of Leeds, Clarendon Way, Leeds, LS2 9NL, UK. Email: w.harrison@leeds.ac.uk. Tel: +44 (0)113 343 4831.

## Word count

Abstract = 200; Main text = 2996

## Abstract

Where performance comparison of healthcare providers is of interest, characteristics of both patients and the health condition of interest must be balanced across providers for a fair comparison. This is unlikely to be feasible within observational data, as patient population characteristics may vary geographically and patient care may vary by characteristics of the health condition. We simulated data for patients and providers, based on a previously utilized real-world dataset, and separately considered both binary and continuous covariate-effects at the upper level. Multilevel latent class (MLC) modelling is proposed to partition a prediction focus at the patient level (accommodating 'casemix') and a causal inference focus at the provider level. The MLC model recovered a range of simulated Trust-level effects. Median recovered values were almost identical to simulated values for the binary Trust-level covariate, and we observed successful recovery of the continuous Trust-level covariate with at least 3 latent Trust classes. Credible intervals widen as the error variance increases. The MLC approach successfully partitioned modelling for prediction and for causal inference, addressing the potential conflict between these two distinct analytical strategies. This improves upon strategies which only adjust for differential selection. Patient-level variation and measurement uncertainty are accommodated within the latent classes.

## Keywords

## Key Messages

- Performance comparison of healthcare providers requires balance of patient characteristics and the health condition across providers, which is improbable due to heterogeneity.

- We use simulated data to show how multilevel latent class (MLC) modelling can accommodate patient predicted 'casemix' and simultaneously evaluate the causal impact of provider-level factors, such as surgeon specialty or available beds.

- We observed a generally successful recovery of simulated provider-level covariate effects, for both binary and continuous factors.

- MLC modelling has the utility to partition different modelling approaches across a hierarchy, and there is much scope for further development.

## Abbreviations

| | |
|---|---|
| 1P | 1 Patient Class |
| 2T, 3T, 4T or 5T | 1, 2, 4 or 5 Trust Classes |
| $\beta_T$ | Trust-level coefficient |
| CI | Credible Interval |
| MLC | Multilevel Latent Class |

**Introduction**

Epidemiologic studies commonly focus on investigating the effect of an exposure on a health outcome at the individual level, while accommodating variation at the upper level of a hierarchy within a multilevel framework. Interest may alternatively lie in the evaluation of differences across an upper level, for instance when seeking to evaluate the performance of healthcare provision, or to assess the effect of provider-level factors on patient outcomes. In this situation, a fair comparison can only follow if individuals' characteristics (such as demographics) and the severity of the health condition ('casemix') are balanced across providers. These circumstances are not very plausible for most aspects of healthcare provision due to inherent heterogeneity in the characteristics of individuals entering the healthcare system, dependent in part upon geography of residence.[1] This patient-level heterogeneity, or 'casemix', leads to differential access to care, a form of differential selection. Further, we expect that patient care will vary by characteristics of the health condition (e.g. stage of diagnosis for cancer, previous case history of myocardial infarction) and features of their healthcare provider (e.g. available specialist, level of post-operative care), all of which likely impact patient-level health outcomes and contribute to any measured performance differences between providers.

Different modelling approaches are required at each level of a hierarchy when accounting for both patient casemix (which involves prediction methodology) and the evaluation of potential causal influences operating at the upper level (which involves causal inference methodology). Strategies that adjust for differential selection (e.g. matching,[2] stratification,[3] regression[3] and propensity score analysis[4,5]) are not readily adaptable to model separately the causal influences operating at the patient and provider levels; they may even introduce bias within such a complex analytical setting.[6-8]

We propose multilevel latent class (MLC) modelling to exploit the inherent hierarchy of patients nested within healthcare providers, by partitioning the prediction focus at the individual level and the causal inference focus at the provider level. This paper extends the approach established in previous work[9] to include putative causal covariates at the provider level (modelled here as National Health Service Trusts) to estimate the causal influence of provider characteristics on patient outcomes while accounting for the differential selection of patient casemix. We use simulated data to demonstrate proof of principle by exploring the extent to which the MLC model can recover simulated provider-level covariate causal effects.

## Methods

**Data simulation**

We simulated data based on a previously utilized health dataset,[9, 10] to reflect real-world data. The simulated data comprised 24,640 patients and 19 National Health Service Trusts. Figure 1 illustrates the overarching simulation approach to the patient and Trust levels.

*[Insert Figure 1 here]*

We first simulated the patient-level covariates age (at diagnosis), sex, and socio-economic status (the Townsend score of material deprivation) using a trivariate covariance matrix informed by real data. Values were drawn randomly from a normal distribution, and sex was categorized as male or female according to the median threshold. Age and socio-economic status were centered on their mean values; standard deviations were defined as per the real dataset: age standard deviation = 11.6, socio-economic status standard deviation = 3.18. Patient-level data were simulated to be homogeneous. Patients were then randomly assigned to Trusts, and Trust sizes were allowed to vary to reflect differing sizes of geographical area.

At the Trust level, we simulated both binary and continuous effects, although we analyzed them separately. These effects represent competing and causally independent features operating at the provider level, for example whether specialist surgeons are available (binary) or the proportion of available beds (continuous). The binary Trust-level covariate was simulated to have parameterized values of -0.5 or +0.5, with some variability introduced by using a random normal distribution with a small standard deviation of 0.01. The continuous Trust-level covariate was simulated to have parameterized values ranging from -0.5 to +0.5. In each case, simulated values were randomly allocated across Trusts and duplication was allowed, to reflect real-world possibilities.

We calculated continuous patient outcomes for each Trust-level covariate, by combining a linear predictor with a normally distributed error term (with mean = 0 and variance calculated as 33%, 50% or 67% of the median variance of the error-free outcome). The linear predictor combined the patient- and Trust-level covariates using the equation:

$$\beta_{0i} + (\beta_{1i} \times age) + (\beta_{2i} \times sex) + (\beta_{3i} \times SES) + (\beta_T \times Trust\text{-}level\ covariate)$$

$\beta_{0i}$ is a constant term at the patient level $i$, $\beta_{1i}$, $\beta_{2i}$ and $\beta_{3i}$ are the effects of the patient-level covariates 'age', 'sex' and 'socio-economic status' respectively, and $\beta_T$ is the coefficient effect of the Trust-level covariate (binary or continuous) as set during simulation. Values of $\beta_{0i}$, $\beta_{1i}$, $\beta_{2i}$ and $\beta_{3i}$ are log odds values taken from previous analysis of the real-life health dataset, with $\beta_{0i}$ = -0.0265, $\beta_{1i}$ = 0.0547, $\beta_{2i}$ = -0.1368 and $\beta_{3i}$ = 0.0527. We investigated 5 coefficient values for each Trust-level covariate to show consistency of recovery from the simulated values. As an informed basis for analysis, we used the absolute effect of sex (0.137) for the binary Trust-level covariate, and the absolute effect of socio-economic status (0.053) for the continuous Trust-level covariate. Table 1 summarizes the $\beta_T$ values used for each Trust-level covariate.

*[Insert Table 1 here]*

In addition to the 5 Trust-level coefficient values ($\beta_T$) and the 3 error variances (33%, 50% or 67%), we used 3 simulation seeds to generate 45 different parameterizations of unique sets of 100 simulated datasets, for each of the binary and continuous Trust-level covariates. Table 2 summarizes the combinations used.

*[Insert Table 2 here]*

**Multilevel Latent Class (MLC) analysis**

In standard latent class analysis,[11, 12] observations are probabilistically assigned to classes with each observation having a probability of belonging to each latent class, and observations fully assigned across all classes. Multilevel latent class (MLC) models[13, 14] are an extension of single-level latent class analysis with observations probabilistically assigned to latent classes at all levels of the hierarchy. Assignment to classes at the lower level is based on similarities in characteristics,[15] leading to homogeneous latent classes at this level, while latent classes at the upper level may be based on either similarities or differences, dependent on model specification and the research question sought. An optimum solution for all classes at all levels is sought simultaneously using maximum likelihood estimation.[11]

While our simulation starts at the patient level and progresses upwards, modelling starts at the Trust level with a latent construct, followed by casemix adjustment. There is no direct overlap of simulation and analysis, allowing for a more robust assessment of the analytical strategy. Patients are grouped into latent classes based on similarities in characteristics, while for our research question, latent Trust classes are determined based on differences in patient characteristics. This approach yields Trust-level latent classes that are heterogeneous with respect to patient characteristics, such that each Trust-level class contains the same mixture of patient classes; Trust classes are effectively casemix 'adjusted'.

We included simulated values of age, sex and socio-economic status in the fixed part of the model at the patient level, as is typical for a prediction model seeking to predict patient outcomes. We model at this level only to accommodate heterogeneity due to differential selection, not to make meaningful interpretation of patient-level covariates. Additional covariates (e.g. stage at diagnosis) may therefore be included, if available, without concern of invoking bias due to the reversal paradox,[16] which would be an issue if we sought causal inference at this level.[17, 18] Patient-level variation is incorporated via a single patient class.

With patient casemix accommodated through standard prediction strategies, and the MLC model constraint that patient-level classes are balanced across Trust-level classes, any residual differences in patient outcomes are due to unmodelled causal effects operating at the Trust level. We simulated these effects, so our interest is in the comparison between simulated and recovered Trust-level covariate coefficient values. As a minimum, 2 Trust classes are required to distinguish outcome differences; more than 2 classes allow for greater flexibility in modelling variations at the Trust level.

**Trust-level coefficient recovery**

We modelled each simulated dataset using the approach described and calculated a weighted mean outcome for each Trust based on probabilistic assignment to Trust class. Values of the Trust-level coefficient ($\beta_T$) were recovered using single-level regression analysis. We repeated this process for all simulated datasets, and calculated medians and credible intervals (CI: 2.5%, 97.5%) over each set of 100 repeated simulation datasets, for each MLC model scenario, simulated Trust-level coefficient parameter value, and error variance.

**Software**

Stata v14.2[19] was used to perform the simulations, collation of results and single-level linear regression analyses. Latent GOLD[20, 21] was used for all latent variable models, which incorporates an adapted expectation-maximization algorithm[13] for maximum likelihood estimation.

# Results

**Binary Trust-level covariate**

Table 3 shows the results of the analyses for the binary Trust-level covariate. Results were consistent across simulation seeds; models contained 1 patient class (1P) and up to 4 Trust classes (4T).

*[Insert Table 3 here]*

For all combinations, the simulated values of the Trust-level coefficient ($\beta_T$) were within credible intervals for each recovered $\beta_T$ value, and results were consistent across different models and error variances. The median recovered $\beta_T$ was almost identical to the simulated $\beta_T$ for all simulated values except the lowest, regardless of error variance or MLC model. In general, as the error variance increases, the credible intervals became gradually wider, as would be expected, but differences were small.

For the lowest simulated $\beta_T$ value of 0.027, the recovered $\beta_T$ value reduces as the error variance increases (from 0.017-0.018 at the 33% error variance to 0.012-0.013 at the 67% error variance), and some $\beta_T$ coefficients could not be recovered due to some simulated datasets yielding the same probability of class membership for each of the 19 Trusts. We hypothesize that, at very small values of $\beta_T$, the noise introduced when simulating data dominates the value of the $\beta_T$ coefficient and the modelling process is unable to divide the Trusts into identifiably different Trust classes. The number of datasets excluded was small (0-1 at the 33% error variance, 0-5 at the 50% error variance, and 4-11 at the 67% error variance), but some bias may have occurred.

Figure 2 shows the results from table 3 plotted by error variance, demonstrating that the line of equality (where recovered $\beta_T$ equals simulated $\beta_T$) lies almost exactly on the data points and is well within the credible intervals. We included all MLC models and made no distinction between the number of Trust classes.

*[Insert Figure 2 here]*

**Continuous Trust-level covariate**

Table 4 shows the results of the analyses for the continuous Trust-level covariate. Results were consistent across simulation seeds; models contained one patient class (1P) and up to five Trust classes (5T), to reflect the gradual improvement seen as the number of Trust classes increase, revealing how model robustness may warrant more Trust classes than deemed most parsimonious (in contrast to typical latent class modelling strategies that may favor fewer classes to minimize model complexity).

*[Insert Table 4 here]*

For all combinations, the median recovered values of the Trust-level coefficient ($\beta_T$) were lower than those simulated, although simulated values were within the credible intervals for each recovered $\beta_T$ value for models with 3 Trust classes or more. Again, credible intervals widen as the error variance increases, but here they also widen as the simulated $\beta_T$ value increases. Estimates were better for smaller values of the error variance, indicating that an increase in simulated error variance may be dominating the value of the $\beta_T$ coefficient such that the modelling process is less able to separate the Trusts into distinct Trust classes. For the larger simulated $\beta_T$ values, estimates improve as the number of Trust classes increase, and we see this pattern for all values of the error variance. We anticipated this relationship, as more Trust classes are required to robustly differentiate differences between values of the continuous Trust-level covariate across Trusts.

For the lowest simulated $\beta_T$ value of 0.027, the median returned value does not differ much across MLC models, and the credible intervals increase only slightly as the error variance increases. There was a similar, though attenuated, pattern seen for the second lowest simulated $\beta_T$ value of 0.053. Again, some datasets were excluded for the same reasons as described for the binary Trust-level covariate, both for the lowest $\beta_T$ value of 0.027 (41-53 at the 33% error variance, 43-57 at the 50% error variance, and 46-61 at the 67% error variance), and the second lowest $\beta_T$ value of 0.053 (0-4 at the 50% error variance, and 2-6 at the 67% error variance). We therefore excluded the lowest simulated value of $\beta_T$ from any further investigation into the relationship between simulated and recovered values, as too few datasets are included to rely on the results observed.

Figures 3-5 show the results from table 4, excluding the lowest value of $\beta_T$ = 0.011, plotted separately by error variance, and showing the gradually improving relationship between the simulated and recovered values of the Trust-level coefficient as the number of Trust classes increase.

*[Insert Figures 3-5 here]*

**Sensitivity analyses**

In real-world situations, Trust sizes might vary, the division of the binary Trust-level covariate (currently ±0.5) might differ more than we have initially simulated, and the continuous Trust-level covariate (currently ranging from -0.5 to +0.5) may be duplicated across Trusts. To assess the sensitivity of our simulations, we amended each of these aspects individually; we found that none of these choices affected model outcomes.

There was also little difference in the recovered values of $\beta_T$ obtained when we increased the number of simulation datasets to 1,000, supporting the use of 100 simulated datasets per combination of simulation parameters.

We considered the implication of simulating 50 Trusts (opposed to 19), making no other concurrent changes to the simulation process. We used the same MLC models but modelled only 50% error variance, to assess changes in recovered estimates without needing to replicate the entire set of results. Recovered estimates were reduced for lower simulated $\beta_T$ values when we considered a binary Trust-level covariate and for all simulated $\beta_T$ values when we considered a continuous Trust-level covariate. Although all simulated $\beta_T$ values (except the lowest) remained within credible intervals of the recovered values for the binary Trust-level covariate, this was no longer the case for the continuous Trust-level covariate, due to a general narrowing of credible intervals throughout. We also modelled with 1 patient-class and 10 Trust-classes for both 19 and 50 Trusts (but considered only 50% error variance and 1 simulation seed to reduce computational requirements). Results did not show improvement compared to the 1 patient-class 5 Trust-class MLC models for either 19 or 50 Trusts.

## Discussion

By adopting the MLC modelling strategy, we have shown a successful recovery of Trust-level coefficient parameter values. Simulated values were within credible intervals of the recovered values throughout for the binary Trust-level covariate, and when the number of Trust classes were at least 3 for the continuous Trust-level covariate. We observed some attenuation of effect (i.e. median estimates were lower than simulated true values) for the continuous Trust-level covariate, which is important when considering larger numbers of Trusts. In this situation, effects seen may be lower than the 'true' effects, particularly for greater error variances and smaller numbers of Trust classes. Whilst simulations with 50 Trusts can support more Trust classes compared to those with 19 Trusts, there is no evidence that increasing the number of Trust classes is a solution.

Lower simulated values of the Trust-level coefficients were not recovered as well as higher values, and datasets were excluded when these coefficients could not be recovered at all. It is possible that the variability introduced in the covariates during simulation dominates the coefficient parameter value such that it is harder to estimate within the modelling process. Bias may also exist due to dataset exclusions. It is reassuring, therefore, that these values remain within credible intervals of the recovered values.

The illustrated MLC approach has several advantages. We have partitioned modelling for prediction (casemix adjustment) and for causal inference (assessing the putative causal impact of Trust characteristics) at different levels of the data structure, thus performing adjustment for differential selection at the patient level while allowing for investigation of causal insights at the Trust level. This serves to overcome the potential conflict between two distinctly separate analytical strategies. Uncertainty surrounding class membership is implicit within the latent classes, since observations may belong to all classes, with probabilities determined empirically; the latent class approach therefore accommodates uncertainty better than standard regression modelling. Unlike other casemix adjustment strategies, this approach accommodates both patient-level variation due to unmeasured covariates, and measurement uncertainty within observed covariates, all within the latent constructs adopted.

There is much scope for extension and further development. For simplification, and proof of principle, we analyzed competing and causally independent binary and continuous Trust-level covariate effects separately, making the assessment of total causal effect straightforward. Multiple covariates can logically be included in combination at the Trust level within a robust causal framework,[22] supported by the construction of a multivariable directed acyclic graph (DAG)[23] to resolve which covariates are required to address separate research questions pertaining to the putative causal impacts of each covariate in the DAG.[24] The same modelling framework may accommodate a binary outcome variable, although the fixed binomial error variance of $\pi^2/3$ at the patient level has implications on the effect of the variance structure at higher levels, which may serve to distort the relationship between simulated and recovered values. Also for simplification, we simulated patient-level data to be homogeneous, and modelled using only 1 patient class. The MLC approach allows for any number of patient classes to be specified, to ensure casemix balance across Trust classes by accommodating patient heterogeneity. More complex simulations with mixtures of patient subgroups can therefore be explored in future evaluation. Additional complexities in casemix (e.g. treatment variables) can also be incorporated, in combination with all other potential complexities – this is not typically considered within standard casemix modelling strategies, yet arguably it should be if we assume that the treatment given should part explain the heterogeneity in patient outcomes that are observed.

# References

1. Office for National Statistics. *Population estimates analysis tool*. [Online]. Available from: https://www.ons.gov.uk/peoplepopulationandcommunity/populationandmigration/populationestimates/datasets/populationestimatesanalysistool. [Accessed 5 February 2019].

2. Rothman K, Greenland S, Lash T. *Modern Epidemiology*. Boston: Little Brown and Co.; 1986.

3. Normand S-LT, Sykora K, Li P, Mamdani M, Rochon PA, Anderson GM. Readers guide to critical appraisal of cohort studies: 3. Analytical strategies to reduce confounding. *BMJ* 2005;330(7498): 1021-1023.

4. Rosenbaum PR, Rubin DB. The central role of the propensity score in observational studies for causal effects. *Biometrika* 1983;70(1):41-55.

5. Rosenbaum PR, Rubin DB. Reducing bias in observational studies using subclassification on the propensity score. *J Am Stat Assoc* 1984;79(387):516-524.

6. Deeks JJ, Dinnes J, D'Amico R, Sowden AJ, Sakarovitch C, Song F, Petticrew M, Altman DG. Evaluating non-randomised intervention studies. *Health Technol Assess* 2003;7(27):iii-x, 1-173.

7. Pearl J. Remarks on the method of propensity score [letter]. *Stat Med* 2009;28(9):1415-1416.

8. Pearl J. Understanding bias amplification [invited commentary]. *Am J Epidemiol* 2011;174(11): 1223-1227.

9. Gilthorpe MS, Harrison WJ, Downing A, Forman D, West RM. Multilevel latent class casemix modelling: a novel approach to accommodate patient casemix. *BMC Health Serv Res* 2011;11(53).

10. Harrison WJ, Gilthorpe MS, Downing A, Baxter PD. Multilevel latent class modelling of colorectal cancer survival status at three years and socioeconomic background whilst incorporating stage of disease. *Int J Stat Probab* 2013;2(3):85-95.

11. Goodman LA. Exploratory latent structure analysis using both identifiable and unidentifiable models. *Biometrika* 1974;61(2):215-231.

12. Magidson J, Vermunt JK. Latent class models. In: Kaplan D, ed. *The Sage handbook of quantitative methodology for the social sciences*. Thousand Oaks, CA: Sage; 2004. p. 175-198.

13. Vermunt JK. Multilevel latent class models. *Sociol Methodol* 2003;33(1):213-239.

14. Vermunt JK. Latent class and finite mixture models for multilevel data sets. *Stat Methods Med Res* 2008;17(1):33-51.

15. Skrondal A, Rabe-Hesketh S. *Generalized latent variable modeling: multilevel, longitudinal, and Structural Equation Models.* Boca Raton, FL: Chapman and Hall/CRC; 2004.

16. Stigler SM. *Statistics on the table: the history of statistical concepts and methods.* Cambridge, MA: Harvard University Press; 1999.

17. Tu Y-K, West R, Ellison GTH, Gilthorpe MS. Why evidence for the fetal origins of adult disease might be a statistical artifact: the "reversal paradox" for the relation between birth weight and blood pressure in later life. *Am J Epidemiol* 2005;161(1):27-32.

18. Pearl J, Glymour M, Jewell NP. *Causal inference in statistics: a primer*. Chichester, UK: Wiley; 2016.

19. StataCorp. *Stata statistical software: release 14*. College Station, TX: StataCorp LP; 2015.

20. Vermunt JK, Magidson J. *Latent GOLD 4.0 user's guide.* Belmont, MA: Statistical Innovations Inc.; 2005.

21. Vermunt JK, Magidson J. *LG-syntax user's guide: manual for Latent GOLD 5.0 syntax module.* Belmont, MA: Statistical Innovations Inc.; 2013.

22. Greenland S, Pearl J, Robins JM. Causal diagrams for epidemiologic research. *Epidemiology* 1999; 10(1):37-48.

23. Pearl J. *Causality: models, reasoning and inference.* Cambridge, UK: Cambridge University Press; 2000.

24. Westreich D, Greenland S. The table 2 fallacy: presenting and interpreting confounder and modifier coefficients. *Am J Epidemiol* 2013;177(4):292-8.

## Tables

**Table 1. Trust-level Coefficient Values for the Binary and Continuous Trust-level Covariates.**

|  | $\beta_T$ Coefficient Values | |
|---|---|---|
| $\beta_T$ Effect | Binary Trust-level Covariate | Continuous Trust-level Covariate |
| One fifth effect | 0.027 | 0.011 |
| Effect of sex or deprivation | 0.137 | 0.053 |
| Additional value | 0.250 | 0.120 |
| Additional value | 0.500 | 0.200 |
| Five times effect | 0.684 | 0.264 |

$\beta_T$ – Trust-level coefficient value

**Table 2. Summary of Combinations Used in Data Simulations**

| $\beta_T$ Coefficient | Error variance | Simulation seeds | Binary Trust-level covariate | Continuous Trust-level covariate |
|---|---|---|---|---|
| One fifth effect | 33% | 1, 2, 3 | 9 sets of 100 | 9 sets of 100 |
|  | 50% | 1, 2, 3 |  |  |
|  | 67% | 1, 2, 3 |  |  |
| Effect of sex or deprivation | 33% | 1, 2, 3 | 9 sets of 100 | 9 sets of 100 |
|  | 50% | 1, 2, 3 |  |  |
|  | 67% | 1, 2, 3 |  |  |
| Additional value | 33% | 1, 2, 3 | 9 sets of 100 | 9 sets of 100 |
|  | 50% | 1, 2, 3 |  |  |
|  | 67% | 1, 2, 3 |  |  |
| Additional value | 33% | 1, 2, 3 | 9 sets of 100 | 9 sets of 100 |
|  | 50% | 1, 2, 3 |  |  |
|  | 67% | 1, 2, 3 |  |  |
| Five times effect | 33% | 1, 2, 3 | 9 sets of 100 | 9 sets of 100 |
|  | 50% | 1, 2, 3 |  |  |
|  | 67% | 1, 2, 3 |  |  |

$\beta_T$ – Trust-level coefficient value

**Table 3. Simulated and Recovered Values of the Trust-level Coefficient for the Binary Trust-level Covariate**

| Simulated $\beta_T$ Coefficient | MLC Model | Recovered $\beta_T$ Coefficient | | | | | |
|---|---|---|---|---|---|---|---|
| | | Error variance 33% | | Error Variance 50% | | Error Variance 67% | |
| | | Median[a] | CI | Median[a] | CI | Median[a] | CI |
| 0.027 | 1P-2T | 0.017 | 0.005, 0.030 | 0.014 | 0.002, 0.029 | 0.012 | 0.002, 0.028 |
| | 1P-3T | 0.018 | 0.005, 0.031 | 0.015 | 0.003, 0.030 | 0.013 | 0.002, 0.029 |
| | 1P-4T | 0.018 | 0.005, 0.031 | 0.015 | 0.003, 0.030 | 0.013 | 0.002, 0.029 |
| 0.137 | 1P-2T | 0.137 | 0.126, 0.146 | 0.136 | 0.123, 0.148 | 0.136 | 0.119, 0.149 |
| | 1P-3T | 0.136 | 0.126, 0.146 | 0.136 | 0.123, 0.148 | 0.136 | 0.118, 0.150 |
| | 1P-4T | 0.136 | 0.126, 0.146 | 0.136 | 0.123, 0.149 | 0.136 | 0.118, 0.150 |
| 0.250 | 1P-2T | 0.250 | 0.239, 0.259 | 0.250 | 0.237, 0.261 | 0.250 | 0.235, 0.263 |
| | 1P-3T | 0.250 | 0.239, 0.259 | 0.250 | 0.237, 0.261 | 0.249 | 0.235, 0.263 |
| | 1P-4T | 0.250 | 0.239, 0.259 | 0.250 | 0.237, 0.261 | 0.249 | 0.235, 0.263 |
| 0.500 | 1P-2T | 0.499 | 0.489, 0.509 | 0.499 | 0.486, 0.511 | 0.499 | 0.484, 0.513 |
| | 1P-3T | 0.499 | 0.489, 0.509 | 0.499 | 0.486, 0.511 | 0.499 | 0.484, 0.512 |
| | 1P-4T | 0.499 | 0.489, 0.509 | 0.499 | 0.486, 0.511 | 0.499 | 0.485, 0.513 |
| 0.684 | 1P-2T | 0.683 | 0.672, 0.693 | 0.683 | 0.670, 0.695 | 0.683 | 0.668, 0.697 |
| | 1P-3T | 0.683 | 0.673, 0.693 | 0.683 | 0.670, 0.695 | 0.683 | 0.668, 0.696 |
| | 1P-4T | 0.683 | 0.673, 0.693 | 0.683 | 0.670, 0.695 | 0.683 | 0.668, 0.697 |

1P – 1 Patient Class; 2T, 3T or 4T – 2, 3 or 4 Trust Classes; $\beta_T$ – Trust-level Coefficient Value; CI – Credible Interval

[a]median averaged over 3 simulation seeds

**Table 4. Simulated and Recovered Values of the Trust-level Coefficient for the Continuous Trust-level Covariate**

| Simulated $\beta_T$ Coefficient | MLC Model | Recovered $\beta_T$ Coefficient | | | | | |
|---|---|---|---|---|---|---|---|
| | | Error variance 33% | | Error Variance 50% | | Error Variance 67% | |
| | | Median[a] | CI | Median[a] | CI | Median[a] | CI |
| 0.011 | 1P-2T | 0.003 | -0.001, 0.013 | 0.003 | -0.002, 0.015 | 0.003 | -0.003, 0.017 |
| | 1P-3T | 0.003 | -0.001, 0.014 | 0.003 | -0.002, 0.015 | 0.003 | -0.002, 0.017 |
| | 1P-4T | 0.003 | -0.001, 0.014 | 0.003 | -0.002, 0.015 | 0.003 | -0.002, 0.017 |
| | 1P-5T | 0.003 | -0.001, 0.014 | 0.003 | -0.002, 0.015 | 0.004 | -0.003, 0.017 |
| 0.053 | 1P-2T | 0.032 | 0.011, 0.051 | 0.029 | 0.008, 0.052 | 0.027 | 0.003, 0.052 |
| | 1P-3T | 0.036 | 0.012, 0.056 | 0.031 | 0.008, 0.055 | 0.028 | 0.004, 0.055 |
| | 1P-4T | 0.036 | 0.012, 0.056 | 0.032 | 0.008, 0.055 | 0.028 | 0.004, 0.054 |
| | 1P-5T | 0.036 | 0.012, 0.056 | 0.031 | 0.007, 0.055 | 0.028 | 0.005, 0.054 |
| 0.120 | 1P-2T | 0.090 | 0.063, 0.113 | 0.087 | 0.058, 0.115 | 0.085 | 0.054, 0.115 |
| | 1P-3T | 0.105 | 0.079, 0.124 | 0.101 | 0.071, 0.126 | 0.098 | 0.062, 0.126 |
| | 1P-4T | 0.107 | 0.084, 0.127 | 0.103 | 0.073, 0.126 | 0.099 | 0.063, 0.127 |
| | 1P-5T | 0.109 | 0.085, 0.127 | 0.104 | 0.073, 0.129 | 0.100 | 0.063, 0.129 |
| 0.200 | 1P-2T | 0.153 | 0.113, 0.184 | 0.152 | 0.111, 0.185 | 0.151 | 0.109, 0.186 |
| | 1P-3T | 0.182 | 0.154, 0.201 | 0.180 | 0.148, 0.203 | 0.178 | 0.143, 0.205 |
| | 1P-4T | 0.188 | 0.165, 0.207 | 0.186 | 0.159, 0.209 | 0.183 | 0.149, 0.209 |
| | 1P-5T | 0.191 | 0.168, 0.210 | 0.188 | 0.161, 0.210 | 0.186 | 0.154, 0.212 |
| 0.264 | 1P-2T | 0.201 | 0.153, 0.241 | 0.202 | 0.149, 0.242 | 0.202 | 0.147, 0.243 |
| | 1P-3T | 0.240 | 0.207, 0.263 | 0.240 | 0.203, 0.264 | 0.238 | 0.202, 0.266 |
| | 1P-4T | 0.250 | 0.223, 0.269 | 0.249 | 0.218, 0.271 | 0.248 | 0.213, 0.274 |
| | 1P-5T | 0.254 | 0.230, 0.274 | 0.252 | 0.225, 0.277 | 0.251 | 0.220, 0.277 |

1P – 1 Patient Class; 2T, 3T, 4T or 5T – 2, 3, 4 or 5 Trust Classes; $\beta_T$ – Trust-level Coefficient Value; CI – Credible Interval

[a]median averaged over 3 simulation seeds

# Figures

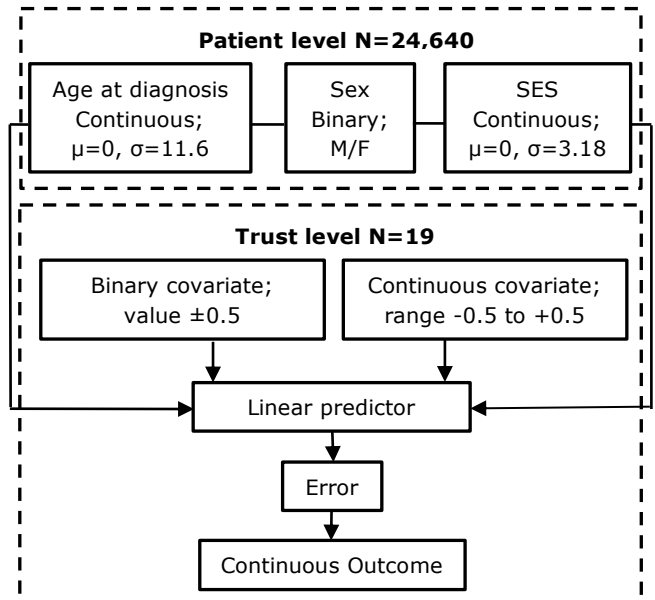

**Figure 1. Overarching simulation approach to the patient and Trust levels**

µ – mean, σ – standard deviation, SES – socio-economic status, N – total number of unique observations at patient or Trust level

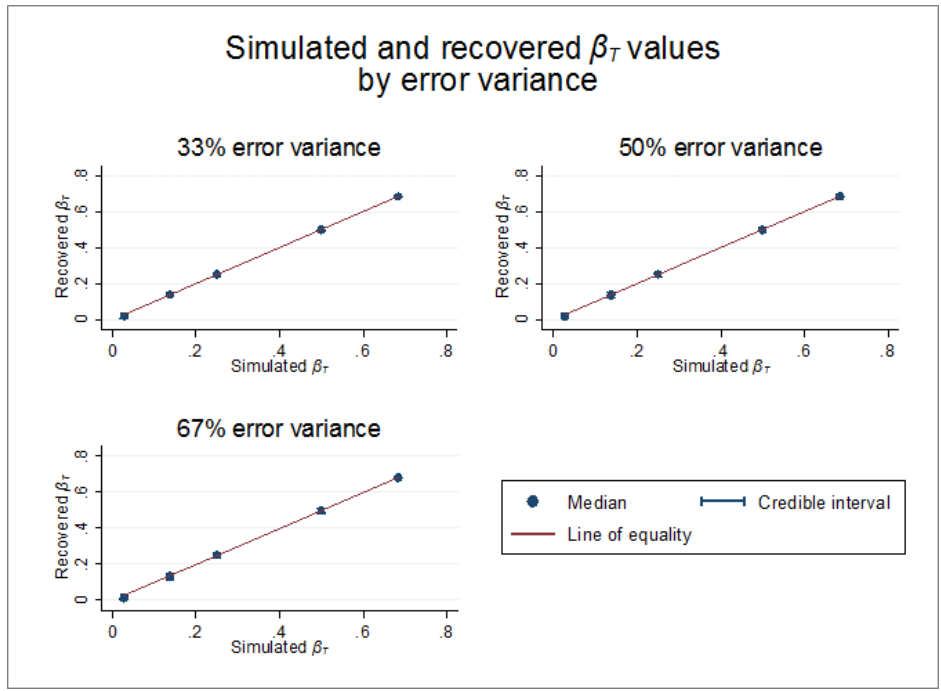

**Figure 2. Plot showing $\beta_T$ relationship for the binary Trust-level covariate**

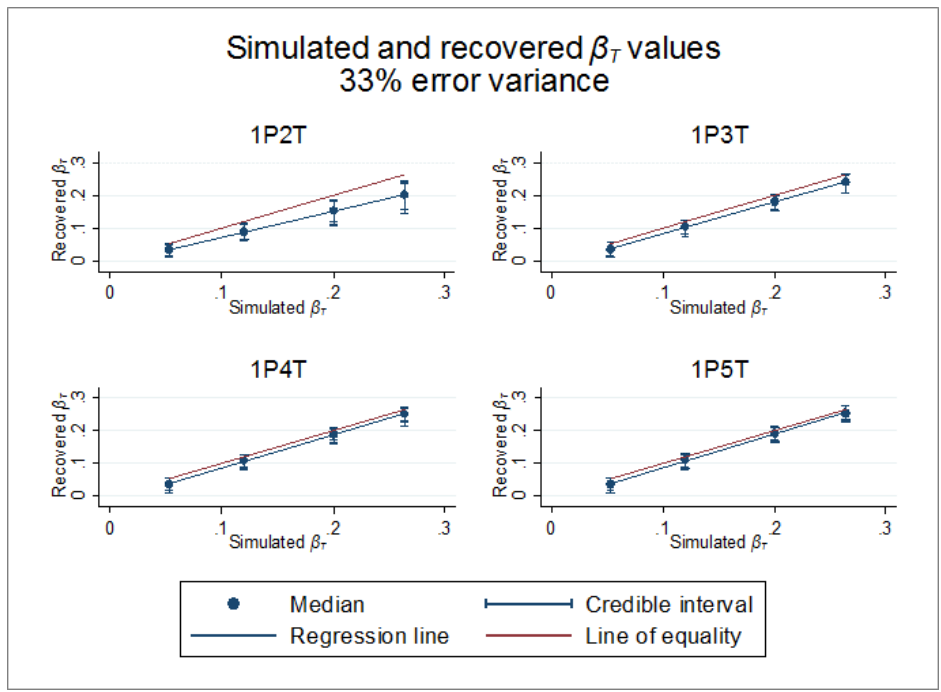

**Figure 3. Plot showing $\beta_T$ relationship for the continuous Trust-level covariate; 33% error variance**

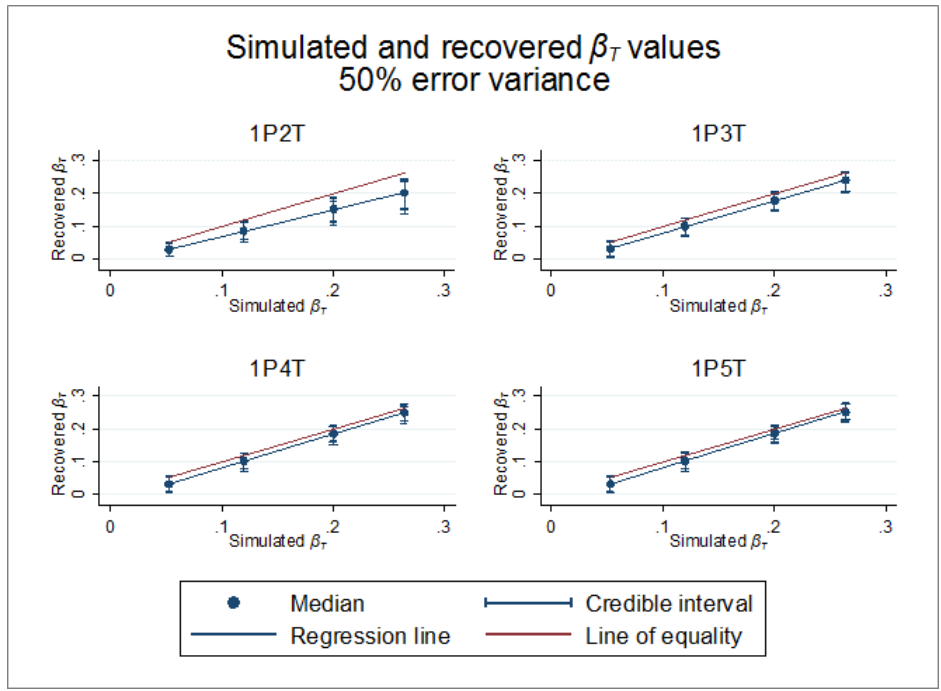

**Figure 4.** Plot showing $\beta_T$ relationship for the continuous Trust-level covariate; 50% error variance

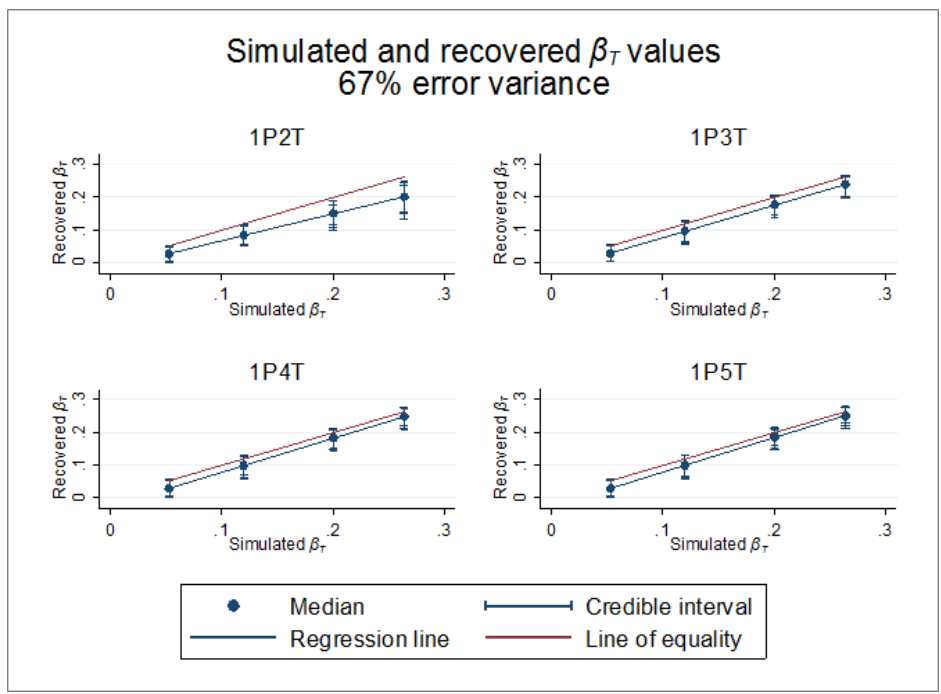

**Figure 5.** Plot showing $\beta_T$ relationship for the continuous Trust-level covariate; 67% error variance

**Funding**

This study received no specific funding; MSG is supported by The Alan Turing Institute [grant number EP/N510129/1].

**Author Contributions**

MSG conceived the idea and with WJH designed the study. WJH developed the idea, simulated the data, performed all analyses, interpreted the simulations and drafted the manuscript. MSG and PDB contributed substantially to data interpretation and critically revised the manuscript. All Authors approved the final version of the manuscript before submission. WJH accepts full responsibility for the work and conduct of the study, had full access to the data and controlled the decision to publish.

**Conflict of Interest**

The Authors declare that there are no conflicts of interest.